\documentclass[aps,prl,reprint,superscriptaddress,floatfix,longbibliography]{revtex4-2}
\usepackage{amsmath}
\usepackage{amssymb}
\usepackage{amsthm}
\usepackage{bm}
\usepackage{graphicx}
\usepackage{algorithm}
\usepackage[noend]{algpseudocode}

\usepackage{siunitx}

\newcommand{\bs}[1]{\bm{#1}}

\newcommand{\dd}{\mathrm{d}}

\begin{document}

\title{Optimal transition in underdamped systems with memory}
\author{Tianyu Luo}
\author{Yunxin Zhang}
\affiliation{School of Mathematical Sciences, Fudan University, Shanghai 200433, CHINA.\\
Shanghai Key Laboratory for Contemporary Applied Mathematics,\\
Laboratory of Mathematics for Nonlinear Science, Fudan University, Shanghai 200433, CHINA.\\
tyluo24@m.fudan.edu.cn\qquad xyz@fudan.edu.cn}

\begin{abstract}
Optimal finite-time control is essential for energy-efficient operation of
nanoscale devices. While existing work has largely focused on transitions
between equilibrium states in overdamped systems, many settings of practical
interest---including nanomechanical resonators,
biomolecular conformational dynamics, and quantum Brownian motion---are
governed by underdamped dynamics where both particle inertia and
frequency-dependent friction (memory) play a non-negligible role.
In this study, we analytically and computationally investigate optimal
transitions between nonequilibrium steady states (NESS) for an underdamped
particle in a moving harmonic trap with general memory kernels. We find that
inertia qualitatively alters optimal control in the presence of memory.
Compared to the overdamped case, underdamped dynamics break the time-reversal
symmetry, making the forward and backward optimal protocols
fundamentally distinct. Across the memory-kernel types examined, the
asymmetry, rather than the detailed form of the kernel, governs the structure
of the optimal strategy. These results offer
a unified framework for optimal control in underdamped systems with memory.
\end{abstract}

\maketitle

\section{Introduction}

The miniaturization of devices down to the micro- and nanoscale has driven
sustained interest in optimal finite-time control, where the goal is to
steer a system between two states with minimal energy expenditure
\cite{a1,a2,a11,a12}. Early work focused on transitions between equilibrium
states, revealing that optimal protocols generically exhibit discontinuous
jumps at the start and end of the control interval and are governed by
fundamental symmetry principles relating forward and time-reversed processes
\cite{a3,a4}. Nonequilibrium steady states (NESS), however, are the rule
rather than the exception in biological and synthetic microscale
systems---from molecular motors and ion pumps to energy-conversion
nanodevices---and efficient cyclic operation depends on the ability to
transition between NESS with low energy cost \cite{a5,a6,a7,a8,a10}.
Unlike equilibrium states, NESS require continuous energy input to maintain,
and transitions between them involve a nontrivial interplay of stored energy,
dissipation, and relaxation.
Recently, Monter et al.\ studied optimal NESS transitions for an overdamped
particle in a viscoelastic fluid, establishing that forward and backward
optimal protocols are equivalent under time reversal \cite{a9}.

A crucial assumption underlying all of the above work---both the
equilibrium and NESS optimal control literature---is the overdamped limit,
where the velocity degree of freedom is instantaneously enslaved to the
force. In many settings of practical importance, however, particle inertia
cannot be neglected. Nanomechanical resonators in vacuum \cite{a16},
optically trapped nanoparticles in low-pressure environments \cite{a17},
biomolecular conformational dynamics \cite{a18}, and quantum Brownian motion
\cite{a19} are all governed by underdamped dynamics in which particle
inertia is non-negligible and frequency-dependent friction (memory) is
ubiquitous. Critically, velocity is an independent dynamical variable
that is odd under time reversal, breaking the forward--backward protocol
symmetry that constrains the overdamped case. The memory kernel varies
qualitatively across these systems---exponential, power-law, and more
complex forms---and a general framework for optimal NESS transitions that
accommodates arbitrary memory kernels has been lacking.

In this study, we develop such a framework by combining Markovian
embedding---which maps non-Markovian dynamics onto a tractable linear
system---with analytical transversality conditions and neural-network
optimization via automatic differentiation. The optimal jump condition
at the protocol endpoint is derived analytically, with memory entering
solely through an effective friction coefficient, and optimal protocols
are obtained numerically across several memory kernel families.
We find that inertia qualitatively alters the structure of optimal
control: the asymmetry between forward and backward protocols, rather
than the detailed form of the memory kernel, governs the optimal strategy.
These results provide a unified framework for energy-efficient control in
underdamped systems with memory.


\section{Underdamped dynamics with memory}
The framework developed below handles memory-free and memoryful systems in a
unified manner: setting $\Gamma(t)=0$ decouples the auxiliary variables
$\bs{w}$ from the tracer dynamics and reduces the effective friction to the
bare friction $\gamma_{\mathrm{eff}}=\gamma$, while all expressions for the
excess work and optimal jump condition remain valid.

\subsection{System and dynamics}

We consider a microscopic particle confined in a harmonic optical potential
$U(x) = 1/2\ k(x - \lambda)^2$, where $k$ is the trap stiffness and
$\lambda(t)$ is the time-dependent trap center serving as the control parameter.
The particle is suspended in a viscoelastic fluid \cite{a15} whose internal degrees of freedom
induce a time-delayed response, i.e., memory effects. When particle mass is not
negligible relative to friction, the dynamics are described by the underdamped
generalized Langevin equation (GLE):
\begin{equation}
\begin{cases}
\dd X_t = V_t \dd t,\\[4pt]
\dd V_t = -U'(X_t)\,\dd t - \gamma V_t\,\dd t \\[2pt]
\qquad\; - \int_0^t \Gamma(t-s) V_s \dd s\,\dd t + \dd Z_t,
\end{cases}
\label{eq:gle}
\end{equation}
where $\gamma$ is the instantaneous friction coefficient, $\Gamma(t)$ is the memory
kernel, and $Z_t$ is a colored noise satisfying the fluctuation-dissipation theorem
$\langle \dd Z_t\,\dd Z_s\rangle = \beta^{-1}[\gamma\delta(t-s) + \Gamma(t-s)]\,\dd t\,\dd s$ with
$\beta = 1/(k_B T)$.

The control protocol $\lambda(t)$ drives the system from $\text{NESS}_i$ (corresponding
to a trap moving at constant velocity $v_i$ for $t<0$) to $\text{NESS}_f$ (trap moving
at $v_f$ for $t\geq t_f$). The transition occurs during $0\leq t\leq t_f$.
For $t\geq t_f$, the protocol is fixed at $\lambda(t) = v_f t + \lambda_0$, allowing
the system to relax to $\text{NESS}_f$.

\subsection{Markovian embedding}

To handle the non-Markovian dynamics, we employ Markovian embedding
\cite{b1,b2,a13}, which introduces $m$ auxiliary variables
$\bs{w} = (w_1,\ldots,w_m)^{\mathrm{T}}$ to absorb the memory. The GLE~\eqref{eq:gle}
is equivalent to the following $(m+2)$-dimensional Markovian system:
\begin{equation}
\begin{cases}
\dd X_t = V_t \dd t, & X(0)=x_0,\\[2pt]
\dd V_t = -\bigl[U'(X_t) + \gamma V_t - \bs{g}^{\mathrm{T}}\bs{w}_t\bigr]\dd t,
          & V(0)=v_0,\\[2pt]
\dd \bs{w}_t = -\bigl[A\bs{w}_t + V_t\,\bs{g}\bigr]\dd t + C\,\dd\bs{B}_t,
          & \bs{w}(0)\sim\mathcal{N}(0,\Sigma),
\end{cases}
\label{eq:markov}
\end{equation}
where $\bs{g}\in\mathbb{R}^m$, and $A,C,\Sigma\in\mathbb{R}^{m\times m}$ are determined
by the memory kernel via
\begin{equation}
\bs{g}^{\mathrm{T}}(sI+A)^{-1}\bs{g} = \mathcal{L}(\Gamma)(s),
\label{eq:laplace_match}
\end{equation}
with $\mathcal{L}(\Gamma)$ the Laplace transform of $\Gamma(t)$.
The fluctuation-dissipation theorem imposes $\Sigma = \beta^{-1}I$ and
$CC^{\mathrm{T}} = -\beta^{-1}(A+A^{\mathrm{T}})$ (see Sec.~B of the Supplemental Material~\cite{SM}
for the full derivation).

In the co-moving frame $\hat{X}_t = X_t - \lambda(t)$, $\hat{V}_t = V_t - v_f$,
the post-protocol dynamics ($t\geq t_f$) become a linear system with constant
coefficients:
\begin{equation}
\dd\bs{Y}_t = M\bs{Y}_t\,\dd t + \bs{b}\,\dd t + N\,\dd\bs{B}_t,
\label{eq:linear_system}
\end{equation}
where $\bs{Y}_t = (\hat{X}_t,\hat{V}_t,\bs{w}_t)^{\mathrm{T}}$ and
\begin{equation}
M = \begin{pmatrix}
0 & 1 & \mathbf{0}^{\mathrm{T}} \\
-k & -\gamma & \bs{g}^{\mathrm{T}} \\
\mathbf{0} & -\bs{g} & -A
\end{pmatrix},\quad
\bs{b} = \begin{pmatrix}
0 \\ -\gamma v_f \\ -v_f\bs{g}
\end{pmatrix},\quad
N = \begin{pmatrix}
0 \\ 0 \\ C
\end{pmatrix}.
\label{eq:M_matrix}
\end{equation}

The final NESS corresponds to a Gaussian distribution with mean
$\bs{\mu} = -M^{-1}\bs{b}$ and covariance satisfying the Lyapunov equation
$M\Sigma + \Sigma M^{\mathrm{T}} + NN^{\mathrm{T}} = 0$.
Explicit calculation yields
\begin{equation}
\bs{\mu} = \begin{pmatrix}
-\frac{v_f}{k}(\gamma + \bs{g}^{\mathrm{T}}A^{-1}\bs{g}) \\[2pt]
0 \\[2pt]
-v_f A^{-1}\bs{g}
\end{pmatrix}.
\label{eq:mu_ness}
\end{equation}

\subsection{Excess work}

The total work performed on the system during a NESS-to-NESS transition includes
a divergent housekeeping contribution required to maintain the steady state.
The physically meaningful cost functional is the \emph{excess work}
\begin{equation}
W_{\mathrm{ex}} = W_{\mathrm{tot}} - \int_{t_i}^{t_f}
\langle J^{W} \rangle_{\lambda(t)}^{\mathrm{ss}}\,\dd t,
\label{eq:excess_def}
\end{equation}
where $\langle J^{W} \rangle_{\lambda(t)}^{\mathrm{ss}}$ is the steady work flow
at fixed $\lambda(t)$.

For a harmonic potential with deterministic control $\lambda(t)$, the expected
excess work simplifies to an expression depending only on the mean trajectory
$\bar{x}(t) = \langle X_t\rangle$ (see Sec.~D of the Supplemental Material~\cite{SM} for proof):
\begin{equation}
W_{\mathrm{ex}} = -k\int_0^\infty \dot{\lambda}(t)\,
\bigl(\bar{x}(t)-\lambda(t)\bigr)\,\dd t
\;-\; \int_{t_f}^\infty P_{\mathrm{hk}}\,\dd t,
\label{eq:excess_mean}
\end{equation}
where the housekeeping power is
\begin{equation}
P_{\mathrm{hk}} = \gamma v_f^2 + v_f^2\,\bs{g}^{\mathrm{T}}A^{-1}\bs{g}.
\label{eq:Phk}
\end{equation}

The excess work naturally splits into two contributions:
$W_{\mathrm{ex}} = W_{\mathrm{proto}} + W_{t\geq t_f}$, where $W_{\mathrm{proto}}$
is the work during the active control phase $[0,t_f]$ and $W_{t\geq t_f}$ is the
relaxation work after the protocol ends.

\section{Results}

\subsection{Optimal jump condition at the protocol endpoint}
\label{sec:results_jump}

A central analytical result is the optimal jump condition at the protocol endpoint
$t=t_f$. This is obtained from the transversality condition of the Euler--Lagrange
equations with free terminal state (full derivation in Sec.~C of the Supplemental Material~\cite{SM}):
\begin{equation}
\left.\frac{\partial\mathcal{L}}{\partial\dot{\lambda}}\right|_{t_f^-}
+ \frac{\partial W_{t\geq t_f}}{\partial\lambda(t_f^+)} = 0,
\label{eq:transversality}
\end{equation}
where $\mathcal{L} = k\dot{\lambda}(\lambda-x)$ is the Lagrangian.
Evaluating this condition yields the \emph{optimal jump}:
\begin{equation}
\Delta\lambda \equiv \lambda(t_f^+) - \lambda(t_f^-)
= -\frac{\gamma + \bs{g}^{\mathrm{T}}A^{-1}\bs{g}}{k}\,v_f.
\label{eq:jump_condition}
\end{equation}

Equivalently, the relative displacement after the jump satisfies
$x(t_f) - \lambda(t_f^+) = -\gamma_{\mathrm{eff}}v_f/k$, which is exactly the
NESS relative displacement.
Equation~\eqref{eq:jump_condition} thus confirms that the optimal protocol must
include an instantaneous jump at $t=t_f$ that places the trap at the position
required for the system to already be in the correct NESS relative configuration,
enabling minimum-energy relaxation to $\text{NESS}_f$.

The jump magnitude depends on the effective friction coefficient
$\gamma_{\mathrm{eff}} = \gamma + \bs{g}^{\mathrm{T}}A^{-1}\bs{g}$, which
incorporates both the instantaneous friction $\gamma$ and the integrated memory
kernel $\int_0^\infty \Gamma(t)\,\dd t = \bs{g}^{\mathrm{T}}A^{-1}\bs{g}$.

\subsection{Numerical optimization via neural networks}
\label{sec:results_nn}

While the jump condition~\eqref{eq:jump_condition} provides the optimal
endpoint behavior, the full optimal protocol $\lambda(t)$ for $0<t<t_f$ must
be determined numerically due to the high dimensionality of the Euler--Lagrange
system ($2m+4$ coupled equations; see Sec.~C of the Supplemental Material~\cite{SM}).

We parameterize the protocol using a feedforward neural network
$\mathcal{N}_\theta: [0,1] \to \mathbb{R}$ that maps normalized time
$\tau = t/t_f$ to the trap position $\lambda(\tau)$. The network weights $\theta$
are optimized to minimize the total excess work $W_{\mathrm{ex}}$, computed by
deterministically integrating the mean dynamics from the initial NESS.
Gradients $\nabla_\theta W_{\mathrm{ex}}$ are evaluated via automatic
differentiation using JAX/Flax \cite{b3,a14}, enabling efficient gradient-based
optimization.

The excess work is computed in three parts. The protocol work
$W_{\mathrm{proto}}$ is integrated along the trajectory during $[0,t_f]$
using the Stratonovich rule. The jump work $W_{\mathrm{jump}}$ accounts for
the instantaneous contribution from the optimal
jump~\eqref{eq:jump_condition} at $t=t_f$. The relaxation work
$W_{\mathrm{relax}}$ is evaluated via the analytical expression
$W_{t\geq t_f} = -k v_f\,[M^{-1}\hat{\bs{Y}}_{t_f}]_1$
(derived in Sec.~D of the Supplemental Material~\cite{SM}), which eliminates the need to
numerically integrate the infinite-time tail in~\eqref{eq:excess_mean}.

\subsection{Exponential-sum memory kernels}
\label{sec:results_exp}

The simplest and most important case is when the memory kernel can be expressed
as a sum of exponentials:
\begin{equation}
\Gamma(t) = \sum_{i=1}^{m} \kappa_i^2 e^{-\alpha_i t},
\label{eq:exp_kernel}
\end{equation}
with $\kappa_i,\alpha_i > 0$. Through~\eqref{eq:laplace_match}, this corresponds to
\begin{equation}
A = \operatorname{Diag}(\alpha_1,\ldots,\alpha_m),\qquad
\bs{g} = (\kappa_1,\ldots,\kappa_m)^{\mathrm{T}}.
\label{eq:exp_Ag}
\end{equation}

We implemented the optimization for a two-mode kernel ($m=2$) with parameters
listed in Table~\ref{tab:exp}.

\begin{table}[t]
\caption{Parameters for the exponential-sum memory kernel ($m=2$).}
\label{tab:exp}
\begin{ruledtabular}
\begin{tabular}{l l l}
Parameter & Value & Unit \\
Mass $m$                      & 1.00   & \si{\micro\gram} \\
Tracer friction $\gamma$      & 0.19   & \si{\micro\newton\second\per\meter} \\
Memory kernel $\alpha_1$      & 1.80   & \si{\per\second} \\
Memory kernel $\alpha_2$      & 0.17   & \si{\per\second} \\
Memory kernel $\kappa_1$      & 7.57e-04 & \si{\micro\newton\per\meter\tothe{1/2}} \\
Memory kernel $\kappa_2$      & 1.74e-04 & \si{\micro\newton\per\meter\tothe{1/2}} \\
\end{tabular}
\end{ruledtabular}
\end{table}

\begin{figure}[t]
\includegraphics[width=\columnwidth]{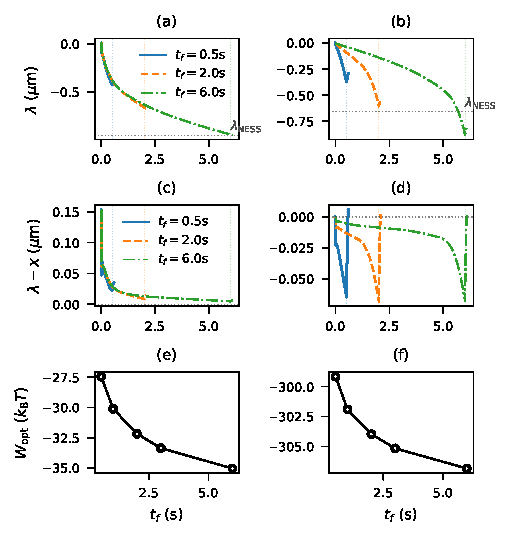}
\caption{Optimal finite-time NESS transition for the exponential-sum kernel
($m=2$; parameters in Table~\ref{tab:exp}).
Left column (a,c,e): deceleration ($v_i=1\to v_f=0$).
Right column (b,d,f): acceleration ($v_i=0\to v_f=1$).
\textbf{(a,b)} Optimal protocol $\lambda(t)$.
\textbf{(c,d)} Trap--particle separation $\lambda(t)-x(t)$.
\textbf{(e,f)} Optimal work $W_{\rm opt}$ versus protocol duration $t_f$.
Forward and reverse protocols are not time-reversal images of each other,
reflecting the broken time-reversal symmetry in the underdamped regime.}
\label{fig:result_exp}
\end{figure}

Figure~\ref{fig:result_exp} shows the optimization results for the
exponential-sum kernel in both directions.
Panels (a,c,e) correspond to deceleration
($v_i=1\,\mu\mathrm{m/s}\to v_f=0$), and
panels (b,d,f) to acceleration
($v_i=0\to v_f=1\,\mu\mathrm{m/s}$).
Several features are noteworthy. First, the optimal
protocol always exhibits the predicted jump at $t=t_f$, consistent with
Eq.~\eqref{eq:jump_condition}. Second, the protocol is non-monotonic
during the transition---overshooting or undershooting the final value to
pre-compensate for the delayed response of the memory modes.
Third, $W_{\rm opt}$ decreases monotonically with $t_f$ in both directions.
The acceleration case ($v_i=0\to v_f=1$) achieves substantially more negative
work than the deceleration case ($v_i=1\to v_f=0$), owing to the large
relaxation contribution from the post-protocol moving NESS.
Most importantly, the optimal protocols for forward and reverse
directions are not time-reversal images of each other
($\lambda^*_{i\to f}\neq\tilde\lambda^*_{f\to i}$), as predicted from
the broken time-reversal symmetry in underdamped dynamics.
This contrasts sharply with the overdamped result where the two coincide.

\subsection{Continued-fraction memory kernels}
\label{sec:results_cf}

A more expressive representation is obtained when the Laplace transform of
the memory kernel admits a continued-fraction expansion:
\begin{equation}
\mathcal{L}(\Gamma)(s) =
\frac{\kappa_1^2}{s+\alpha_1+\frac{\kappa_2^2}{s+\alpha_2+\ddots}},
\label{eq:cf_laplace}
\end{equation}
which corresponds to a tridiagonal matrix $A$:
\begin{equation}
A = \begin{pmatrix}
\alpha_1 & -\kappa_2 & 0 & \cdots & 0 \\
\kappa_2 & \alpha_2 & -\kappa_3 & \cdots & 0 \\
0 & \kappa_3 & \alpha_3 & \cdots & 0 \\
\vdots & \vdots & \vdots & \ddots & -\kappa_m \\
0 & 0 & 0 & \kappa_m & \alpha_m
\end{pmatrix},
\;\;
\bs{g} = (\kappa_1,0,\ldots,0)^{\mathrm{T}},
\label{eq:cf_Ag}
\end{equation}
with $C = \operatorname{Diag}(\sqrt{2\alpha_1},\ldots,\sqrt{2\alpha_m})$
(see Sec.~F of the Supplemental Material~\cite{SM} for details).

This representation is particularly powerful because it can approximate
power-law memory kernels $\Gamma(t) \sim t^{-\gamma}$ ($0<\gamma<1$)
through appropriate choice of the continued-fraction coefficients.
For $\gamma = 1/2$, an exact continued-fraction representation exists,
leading to explicit $A$ and $\bs{g}$ matrices (Sec.~F of the Supplemental Material~\cite{SM}).

\begin{table}[t]
\caption{Parameters for the continued-fraction memory kernel.}
\label{tab:cf}
\begin{ruledtabular}
\begin{tabular}{l S[table-format=2.2] l}
Parameter & {Value} & {Unit} \\
Mass $m$                         & 0.01 & \si{\micro\gram} \\
Trap stiffness $k$               & 4.48 & \si{\micro\newton\per\meter} \\
Tracer friction $\gamma$         & 0.23 & \si{\micro\newton\second\per\meter} \\
Bath 1 stiffness $\kappa_{b1}$   & 0.90 & \si{\micro\newton\per\meter} \\
Bath 1 friction $\gamma_{b1}$    & 0.55 & \si{\micro\newton\second\per\meter} \\
Bath 2 stiffness $\kappa_{b2}$   & 0.04 & \si{\micro\newton\per\meter} \\
Bath 2 friction $\gamma_{b2}$    & 0.28 & \si{\micro\newton\second\per\meter} \\[4pt]
Relaxation time $\tau_1 = \gamma_{b1}/\kappa_{b1}$ & 0.61 & \si{\second} \\
Relaxation time $\tau_2 = \gamma_{b2}/\kappa_{b2}$ & 7.0  & \si{\second} \\
\end{tabular}
\end{ruledtabular}
\end{table}

The continued-fraction representation maps a viscoelastic fluid with multiple
relaxation timescales onto the auxiliary-variable framework, where each rung
of the continued fraction corresponds to an additional internal degree of freedom.
The separation of timescales ($\tau_1 = \SI{0.61}{\second}$,
$\tau_2 = \SI{7.0}{\second}$) in Table~\ref{tab:cf} corresponds to two
viscoelastic baths coupled in series.

\begin{figure}[t]
\includegraphics[width=\columnwidth]{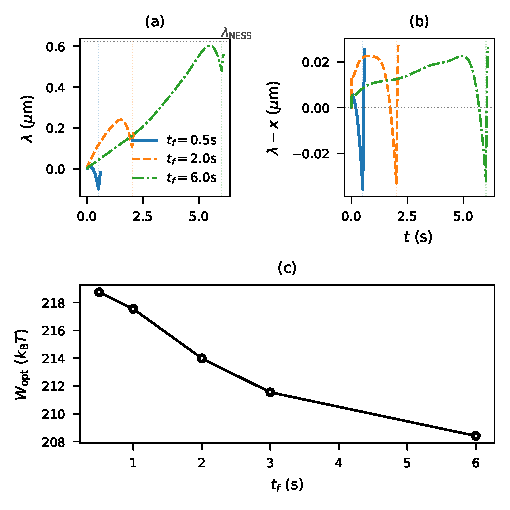}
\caption{Optimal finite-time NESS transition for the continued-fraction kernel
($m=2$; parameters in Table~\ref{tab:cf}).
\textbf{(a)} Optimal protocol $\lambda(t)$.
\textbf{(b)} Trap--particle separation $\lambda(t)-x(t)$.
\textbf{(c)} Optimal work $W_{\rm opt}$ versus protocol duration $t_f$.}
\label{fig:result_cf}
\end{figure}

Figure~\ref{fig:result_cf} shows the optimization results for the
continued-fraction memory kernel with $v_i=0$, $v_f=1\,\mu\mathrm{m/s}$
(acceleration from rest). Compared to the exponential-sum case
(Figure~\ref{fig:result_exp}), several differences are apparent. The continued-fraction
coupling induces a more pronounced protocol overshoot, as the system must
pre-compensate for the serial coupling between the two viscoelastic modes.
The slow bath variable $w_2$, which couples only indirectly through the
tridiagonal structure of $A$, exhibits a prolonged transient on the order of
$\tau_2$, indicating that energy stored in this mode dissipates over the
entire post-protocol relaxation period. The $W_{\rm opt}$--$t_f$ curve
(panel c) confirms that the optimal protocol reduces the work input
substantially relative to the constant-velocity baseline.

\subsection{Power-law memory kernel ($t^{-1/2}$)}
\label{sec:results_pl}

The continued-fraction representation can approximate power-law memory kernels
$\Gamma(t)\propto t^{-\gamma}$ with $\gamma\in(0,1)$. For $\gamma=1/2$, an exact
continued-fraction expansion exists (see Sec.~F of the Supplemental Material~\cite{SM}), yielding
\begin{equation}
\Gamma(t) = \frac{g_1^2}{\sqrt{\pi t}},
\qquad g_1 = 1.96\times10^{-4},
\label{eq:pl_kernel}
\end{equation}
with a sparse $\bs{g}$ vector (only $w_1$ couples to velocity) and a matrix $A$ whose entries
follow a regular pattern. The prefactor is fixed by the requirement that
$\gamma_{\rm eff}=7.30\times10^{-7}$\,Ns/m is independent of the truncation order $m$.

\begin{figure}[t]
\includegraphics[width=\columnwidth]{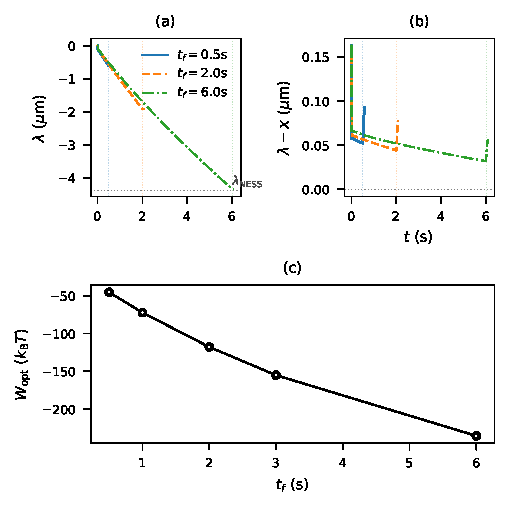}
\caption{Optimal finite-time NESS transition for the power-law $t^{-1/2}$
memory kernel ($m=4$, continued-fraction truncation).
\textbf{(a)} Optimal protocol $\lambda(t)$.
\textbf{(b)} Trap--particle separation $\lambda(t)-x(t)$.
\textbf{(c)} Optimal work $W_{\rm opt}$ versus protocol duration $t_f$.}
\label{fig:result_pl}
\end{figure}

Figure~\ref{fig:result_pl} shows the optimization result for the $m=4$
power-law kernel with $v_i=1\,\mu\mathrm{m/s}$, $v_f=0$
(deceleration to rest). The $W_{\rm opt}$--$t_f$ trend confirms the physical picture:
longer protocols allow the controlled degree of freedom to exchange more energy
with the slowly-relaxing bath modes before they dissipate, reducing the net
work cost. The finite-$m$ truncation preserves passivity ($A\succ 0$, $\Gamma(t)\geq 0$,
$\Re[\hat{\Gamma}(i\omega)]\geq 0$), guaranteeing physical consistency.

\subsection{Overdamped limit and comparison}
\label{sec:results_over}

When particle mass is negligible compared to friction, the system reduces to
the overdamped GLE with memory:
\begin{equation}
\gamma\,\dd X_t + \int_{-\infty}^t \Gamma(t-s)\dot{X}_s\,\dd s\,\dd t
= -U'(X_t)\,\dd t + \dd Z_t.
\label{eq:overdamped}
\end{equation}

The overdamped Markovian embedding (Sec.~E of the Supplemental Material~\cite{SM}) yields an
$(m+1)$-dimensional system with a simpler structure. The excess work during
relaxation admits the compact analytical form
$W_{\mathrm{ex}}^{t>t_f} = k v_f[\mathbf{A}^{-1}\bs{y}(t_f)]_1$,
where $\mathbf{A}$ is the $(m+1)\times(m+1)$ coefficient matrix.

Comparing the underdamped and overdamped results, we find that the excess work
in underdamped systems is systematically larger. This difference arises because
the particle inertia excites additional oscillatory modes in the
particle--memory--bath system, which must subsequently dissipate. In the
overdamped limit, these inertial modes are absent, and the optimal protocol
reduces to the form found by Monter et al.\ \cite{a9}, providing a consistency
check on our framework.

The optimal jump condition~\eqref{eq:jump_condition} also reduces correctly:
in the overdamped limit, the effective friction
$\gamma_{\mathrm{eff}} = \gamma + \bs{g}^{\mathrm{T}}A^{-1}\bs{g}$ replaces
the bare friction, and the jump $\Delta\lambda = -\gamma_{\mathrm{eff}}v_f/k$
matches the known overdamped result.

\section{Discussion and Conclusion}

We have investigated optimal finite-time transitions between NESS for
underdamped systems with memory. The optimal protocol exhibits universal
jumps at the endpoints, given by
$\Delta\lambda = -\gamma_{\mathrm{eff}}v_f/k$, and is generically
non-monotonic during the transition to pre-compensate for memory-induced
delays. The protocol shape varies with the kernel type---from smooth
overshoots for exponential kernels to more pronounced transients for
continued-fraction couplings---but the core structure persists across
all families examined.

Most importantly, underdamped dynamics break the time-reversal symmetry
that constrains the overdamped case: $\lambda^*_{i\to f} \neq
\tilde{\lambda}^*_{f\to i}$ whenever particle inertia is non-negligible.
The asymmetry, rather than the detailed form of the memory kernel,
governs the optimal strategy. This establishes that inertia qualitatively
alters the symmetry class of the optimal control problem.

The Markovian embedding framework combined with neural-network
optimization provides a general tool applicable to arbitrary memory
kernels. Extensions to nonlinear potentials, interacting systems, and
nanoscale engine design are natural next steps.

\begin{acknowledgments}
This study is supported by National Key R\&D Program of China
(2024YFA1012401), the Science and Technology Commission of Shanghai
Municipality (23JC1400501), and Natural Science Foundation of China
(12241103). The authors declare no competing interests.
\end{acknowledgments}

All data and code supporting the findings of this study are available
from the corresponding author upon reasonable request.

\bibliography{sample}

@article{a1,
  author  = {Schmiedl, Tim and Seifert, Udo},
  title   = {Optimal finite-time processes in stochastic thermodynamics},
  journal = {Phys. Rev. Lett.},
  volume  = {98},
  pages   = {108301},
  year    = {2007},
  doi     = {10.1103/PhysRevLett.98.108301},
}

@article{a2,
  author  = {Seifert, Udo},
  title   = {Stochastic thermodynamics, fluctuation theorems and molecular machines},
  journal = {Rep. Prog. Phys.},
  volume  = {75},
  pages   = {126001},
  year    = {2012},
  doi     = {10.1088/0034-4885/75/12/126001},
}

@article{a3,
  author  = {Aurell, Erik and Mej\'{i}a-Monasterio, Carlos and Muratore-Ginanneschi, Paolo},
  title   = {Optimal protocols and optimal transport in stochastic thermodynamics},
  journal = {Phys. Rev. Lett.},
  volume  = {106},
  pages   = {250601},
  year    = {2011},
  doi     = {10.1103/PhysRevLett.106.250601},
}

@article{a4,
  author  = {Loos, Sarah A. M. and Monter, Samuel and Ginot, F\'{e}lix and Bechinger, Clemens},
  title   = {Universal symmetry of optimal control at the microscale},
  journal = {Phys. Rev. X},
  volume  = {14},
  pages   = {021032},
  year    = {2024},
  doi     = {10.1103/PhysRevX.14.021032},
}

@article{a5,
  author  = {Toyabe, Shoichi and Sagawa, Takahiro and Ueda, Masahito and Muneyuki, Eiro and Sano, Masaki},
  title   = {Experimental demonstration of information-to-energy conversion and validation of the generalized {J}arzynski equality},
  journal = {Nat. Phys.},
  volume  = {6},
  pages   = {988--992},
  year    = {2010},
  doi     = {10.1038/nphys1821},
}

@article{a6,
  author  = {J\"{u}licher, Frank and Ajdari, Armand and Prost, Jacques},
  title   = {Modeling molecular motors},
  journal = {Rev. Mod. Phys.},
  volume  = {69},
  pages   = {1269--1282},
  year    = {1997},
  doi     = {10.1103/RevModPhys.69.1269},
}

@article{a9,
  author  = {Monter, Samuel and Loos, Sarah A. M. and Bechinger, Clemens},
  title   = {Optimal transitions between nonequilibrium steady states},
  journal = {Proc. Natl. Acad. Sci. USA},
  volume  = {122},
  pages   = {e2510654122},
  year    = {2025},
  doi     = {10.1073/pnas.2510654122},
}

@article{b1,
  author  = {Kupferman, Raz},
  title   = {Fractional kinetics in {K}ac--{Z}wanzig heat bath models},
  journal = {J. Stat. Phys.},
  volume  = {114},
  pages   = {291--326},
  year    = {2004},
  doi     = {10.1023/B:JOSS.0000003113.22621.f0},
}

@book{b2,
  author    = {Zwanzig, Robert},
  title     = {Nonequilibrium Statistical Mechanics},
  publisher = {Oxford University Press},
  address   = {New York},
  year      = {2001},
  isbn      = {978-0-19-514018-7},
}

@misc{b3,
  author  = {Bradbury, James and Frostig, Roy and Hawkins, Peter and Johnson, Matthew James and Leary, Chris and Maclaurin, Dougal and Necula, George and Paszke, Adam and VanderPlas, Jake and Wanderman-Milne, Skye and Zhang, Qiao},
  title   = {{JAX}: composable transformations of {P}ython+{NumPy} programs},
  url     = {https://github.com/jax-ml/jax},
  year    = {2018},
}

@article{a7,
  author  = {Hatano, Takahiro and Sasa, Shin-ichi},
  title   = {Steady-state thermodynamics of {L}angevin systems},
  journal = {Phys. Rev. Lett.},
  volume  = {86},
  pages   = {3463--3466},
  year    = {2001},
  doi     = {10.1103/PhysRevLett.86.3463},
}

@article{a8,
  author  = {Oono, Yoshitsugu and Paniconi, Marco},
  title   = {Steady state thermodynamics},
  journal = {Prog. Theor. Phys. Suppl.},
  volume  = {130},
  pages   = {29--44},
  year    = {1998},
  doi     = {10.1143/PTPS.130.29},
}

@article{a10,
  author  = {Zulkowski, Patrick R. and Sivak, David A. and DeWeese, Michael R.},
  title   = {Optimal control of transitions between nonequilibrium steady states},
  journal = {PLoS ONE},
  volume  = {8},
  pages   = {e82754},
  year    = {2013},
  doi     = {10.1371/journal.pone.0082754},
}

@article{a11,
  author  = {Bechhoefer, John},
  title   = {Feedback for physicists: {A} tutorial essay on control},
  journal = {Rev. Mod. Phys.},
  volume  = {77},
  pages   = {783--836},
  year    = {2005},
  doi     = {10.1103/RevModPhys.77.783},
}

@article{a12,
  author  = {Jarzynski, Christopher},
  title   = {Nonequilibrium equality for free energy differences},
  journal = {Phys. Rev. Lett.},
  volume  = {78},
  pages   = {2690--2693},
  year    = {1997},
  doi     = {10.1103/PhysRevLett.78.2690},
}

@article{a13,
  author  = {Mori, Hazime},
  title   = {A continued-fraction representation of the time-correlation function},
  journal = {Prog. Theor. Phys.},
  volume  = {34},
  pages   = {399--416},
  year    = {1965},
  doi     = {10.1143/PTP.34.399},
}

@article{a14,
  author  = {Engel, Megan C. and Smith, Jamie A. and Brenner, Michael P.},
  title   = {Optimal control of nonequilibrium systems through automatic differentiation},
  journal = {Phys. Rev. X},
  volume  = {13},
  pages   = {041032},
  year    = {2023},
  doi     = {10.1103/PhysRevX.13.041032},
}

@article{a15,
  author  = {Ginot, F\'{e}lix and Caspers, Jan and Kr\"{u}ger, Matthias and Bechinger, Clemens},
  title   = {Recoil experiments determine the eigenmodes of viscoelastic fluids},
  journal = {New J. Phys.},
  volume  = {24},
  pages   = {123013},
  year    = {2022},
  doi     = {10.1088/1367-2630/aca8d5},
}

@article{a16,
  author  = {Ekinci, Kamil L. and Roukes, Michael L.},
  title   = {Nanoelectromechanical systems},
  journal = {Rev. Sci. Instrum.},
  volume  = {76},
  pages   = {061101},
  year    = {2005},
  doi     = {10.1063/1.1927327},
}

@article{a17,
  author  = {Gieseler, Jan and Deutsch, Bradley and Quidant, Romain and Novotn\'{y}, Luk\'{a}\v{s}},
  title   = {Subkelvin parametric feedback cooling of a laser-trapped nanoparticle},
  journal = {Phys. Rev. Lett.},
  volume  = {109},
  pages   = {103603},
  year    = {2012},
  doi     = {10.1103/PhysRevLett.109.103603},
}

@article{a18,
  author  = {H\"{a}nggi, Peter and Talkner, Peter and Borkovec, Michal},
  title   = {Reaction-rate theory: fifty years after {K}ramers},
  journal = {Rev. Mod. Phys.},
  volume  = {62},
  pages   = {251--341},
  year    = {1990},
  doi     = {10.1103/RevModPhys.62.251},
}

@article{a19,
  author  = {Caldeira, A. O. and Leggett, A. J.},
  title   = {Quantum tunnelling in a dissipative system},
  journal = {Ann. Phys.},
  volume  = {149},
  pages   = {374--456},
  year    = {1983},
  doi     = {10.1016/0003-4916(83)90202-6},
}

@misc{SM,
  year    = {2026},
  note = {See Supplemental Material at [URL will be inserted by publisher] for
          details of the numerical optimization procedure, Markovian embedding,
          Euler--Lagrange derivation, excess work proofs, overdamped limit,
          and continued-fraction representation.},
}

\end{document}